\documentclass[11pt]{article}

\usepackage{latexsym}
\usepackage{color,bm}
\usepackage{amsmath,amssymb,amsthm}
\usepackage{harvard}
\usepackage{charter}

\newcommand{\R}{\mathbb{R}}
\newcommand{\Q}{\mathbb{Q}}
\newcommand{\N}{\mathbb{N}}
\renewcommand{\P}{\mathbb{P}}
\newcommand{\E}{\mathbb{E}}

\newcommand{\cF}{{\cal F}}

\newcommand{\B}{\mathcal{B}}
\newcommand{\F}{\mathcal{F}}
\renewcommand{\leq}{\leqslant}
\renewcommand{\geq}{\geqslant}
\newcommand{\ps}[2]{ \langle #1 , #2 \rangle }
\newcommand{\1}{\mathbf{1}}
\newcommand{\norm}[1]{\left\| #1 \right\|}
\renewcommand{\leq}{\leqslant} \renewcommand{\geq}{\geqslant}
\newcommand{\as}{\mathrm{a.s.}}

\newcommand{\et}{\quad \mathrm{and} \quad}

\renewcommand{\cite}{\citeasnoun}
\newcommand{\rref}[1]{{\rm (\ref{#1})}}

\newcommand{\condexp}[2][\cF_t]{ \E \left[ \left.#2 \right| #1 \right] }

\newcommand{\normstrong}[1]{ \left\| #1 \right\|_{\operatorname{s}}}
\newcommand{\normtv}[1]{ \left\| #1 \right\|_{\operatorname{tot}}}

{\theoremstyle{definition}
\newtheorem{theorem}{Theorem}[section]

\newtheorem{lemma}{Lemma}[section]

\newtheorem{definition}{Definition}[section]
\newtheorem{claim}{Claim}[section]
\newtheorem*{assumption}{Assumption}
}

{\theoremstyle{remark}

\newtheorem{remark}{Remark}[section]

}

\numberwithin{equation}{section}

\begin{document}

\title{On Equilibrium Prices in Continuous Time}

\author{V.~Filipe Martins-da-Rocha \thanks{Graduate School of Economics, Getulio Vargas Foundation, Praia de Botafogo 190, 22.250-900 Rio de Janeiro, Brazil. Email: victor.rocha@fgv.br} \and Frank Riedel \thanks{Institute of Mathematical Economics, Bielefeld University, Postfach 100131 33501 Bielefeld, Germany. Email: friedel@wiwi.uni-bielefeld.de}
}

\maketitle

\begin{abstract}
We combine general equilibrium theory and \emph{th\'{e}orie g\'{e}n\'{e}rale} of stochastic processes to derive structural results about equilibrium state prices.
\medskip


\medskip

\end{abstract}

\section*{Introduction}

 Rational asset prices are the expected present value of future cash flows, properly discounted with a state price, also known as  stochastic discount factor, or Arrow price.
 In economic terms, the state price $\psi_t(\omega)$ is the price of one unit of the num\'{e}raire consumption good (or money) delivered at time $t$ when state $\omega$ prevails.
 With complete markets, the price of a financial asset is  obtained by valuing the asset's payment stream with the state price.
 State-prices determine asset prices, interest rates, and the pricing or equivalent martingale measure.

It is thus important to understand what economic theory can say about the properties of such state prices.
One might fear that the answer is: Not much, in general.
When we take a classical setup where consumption plans come from some $L^p$--space, and preferences are norm-- or Mackey--continuous on the commodity space, the famous existence theorems of general equilibrium theory return state prices in the dual space.\footnote{See the overview by \cite{MasColellZame91} for an account of infinite--dimensional general equilibrium theory. A classic in this regard is the first part of the existence proof in \cite{DuffieZame89}. }
In this generality, a state price is just a nonnegative, measurable, adapted process that satisfies some degree of integrability.
In particular, from  general equilibrium theory, we do not get continuous sample paths, or state prices that are diffusions (as one would like to have to justify a Samuelson--Black--Scholes--type model of asset markets).
One can then impose additional assumptions on preferences and endowments, of course.
\cite{DuffieZame89}, e.g., assume that endowments are It\^{o} processes, and that agents have time--additive smooth expected utility functions. From the first--order conditions of utility maximization, a representative agent argument, and It\^{o}'s lemma, Duffie and Zame obtain equilibrium state-prices that are  diffusions.

One wonders if the structure of time and uncertainty does not allow to derive more structure for state prices in complete generality.
\cite{HindyHuangKreps92} (HHK) try to develop such an approach.
They challenge the implicit assumption that preferences are  norm-- or Mackey--continuous asking when  a rational agent should consider contingent consumption plans as close.
A rational agent should treat small up or downward shifts as close, of course; but she should also react smoothly to  small shifts of (lifetime) consumption plans  over time.
After all, most agents do not care much about getting their retirement payments in twenty five years from now or twenty --five years plus one day.
Technically, one uses the weak topology for distribution functions on the time axis and some kind of $L^p$--topology for uncertainty.
HHK characterize the dual space (where prices come from).
Elements of the dual are given by state prices that are the sum of a martingale and some absolutely continuous process.
HHK's work thus shows that a suitable notion of continuity and the structure of time and uncertainty allow to derive results for state prices.
Unfortunately, there are no equilibria with prices from the HHK dual in general as these spaces are not lattices.
\cite{BankRiedel02} and \cite{Martinsda} establish existence of equilibria in bigger price spaces where the equilibrium price functional is not necessarily continuous on the whole commodity space, but only on the consumption set (the positive cone of the commodity space).

In the present paper we take the latter result as a starting point and ask: what are the positive, linear functionals that are also continuous on the consumption set?
We fully characterize the corresponding state prices, and show that under minimal continuity requirements on preferences and a suitable properness condition,\footnote{Such properness is necessary in infinite--dimensional models, see \cite{MasColellZame91} for a discussion.} equilibria with such state prices exist.

The method to derive the characterization of state prices relies on the \emph{th\'{e}orie g\'{e}n\'{e}rale} of stochastic processes as developed in \cite{DellacherieMeyer75ChpI-IV}.
It turns out that much in the same way as It\^{o}'s theory of stochastic integration is taylor--made for the Samuelson--Black--Scholes theory of asset markets, the
\emph{th\'{e}orie g\'{e}n\'{e}rale} suits our general theory for equilibrium state prices.
This is the first connection of general equilibrium theory and \emph{th\'{e}orie g\'{e}n\'{e}rale}, and we hope that more interesting results can spring from this relation in the future.

For the connoisseurs, we sketch parts of our representation theorem here.
Details concerning notation, if not obvious, are explained in Section \ref{SecModel} below.
We take a nonnegative, linear price functional on the space of all optional random measures with  total variation in $L^p$.
Fixing a stopping time $\tau$, we consider only the restriction on payment streams that pay off at time $\tau$.
This gives us a family $\left(\pi^\tau\right)_{\tau \ \mbox{\tiny stopping time}}$ of linear mappings from $L^p\left(\cF_\tau\right)$ to the real numbers.
With a fixed maturity, there are no issues of shifting etc., so that this mapping is norm--continuous;  Riesz' theorem gives us a random variable $z^\tau \in L^q\left(\cF_\tau\right)$  that represents this mapping.
So we obtain a family of random variables $\left(z^\tau\right)_{\tau \ \mbox{\tiny stopping time}}$ where every $z^\tau$  is $\cF_\tau$--measurable.
We show that this large family is consistent in the sense that we have $$z^\tau=z^\sigma$$ on the event $\{\tau=\sigma\}$ for two stopping times $\sigma$ and $\tau$.
Such families are called $\mathcal{T}$--systems in the  \emph{th\'{e}orie g\'{e}n\'{e}rale}.
The question is if we can find an adapted stochastic process $\left(\psi_t\right)$ such that $\psi_\tau=z^\tau$ for all stopping times $\tau$.
In this case, one says  that $\left(\psi_t\right)$ recollects  the family $\left(z^\tau\right)_{\tau \ \mbox{\tiny stopping time}}$.
It has been shown by \cite{DellacherieLenglart80} that a $\mathcal{T}$--system can be recollected if it is left--continuous in expectation\footnote{This is not sample--path left--continuity. A $\mathcal{T}$--system $\left(w^\tau\right)_{\tau \ \mbox{\tiny stopping time}}$ is left--continuous in expectation if $\E w^{\tau_n} \to \E w^\tau$ whenever $\tau_n \uparrow \tau$ a.s.}(and this kind of continuity is necessary, in general).
Fortunately, the intertemporal topology we use here gives us even continuity in expectation.
Note that we use here the shifting property of the intertemporal topology proposed by HHK: as small shifts over time are considered as close in that topology, the price of a unit of the consumption good delivered at $\tau_n$ has to approach the price of one unit of the consumption good at $\tau$ if $\tau_n \to \tau$.
The recollecting process $(\psi_t)$ is our desired state price.
From continuity in expectation, it is  even cadlag:\footnote{This is again a classic result from the  \emph{th\'{e}orie g\'{e}n\'{e}rale}, see \cite{DellacherieMeyer75ChpI-IV}, Theorem 48.} it has right--continuous sample paths with left hand limits.
It can thus jump.
This might seem puzzling because the dual space for measures on the time axis consists of continuous functions.
And here comes the final clue: the cadlag process  $\left(\psi_t\right)$ is the optional projection of a \emph{continuous, but not necessarily adapted} stochastic process $\left(\xi_t\right)$; this is a result by \cite{Bismut78} and \cite{Emery78}.
Possible jumps in the state price density thus come from the gradual release of information under uncertainty.
To give an example, one might have $\xi_t=Z$ for an $\cF_T$--measurable random variable so that the process $\xi_t$ is constant in time, but not adapted.
The optional projection is then the martingale $\psi_t=\condexp{Z}$. It is well known that in general, martingales jump when information surprises occur.

The paper is organized as follows.
The next section sets up the intertemporal model.
Section~\ref{section:compatible} contains our main theorem characterizing the state prices.
Section~\ref{section:equilibrium} establishes existence of equilibria with such prices and contains examples.

\section{Model and Notation}\label{SecModel}

We consider a stochastic pure exchange economy where a finite set $I$ of agents live in a world of uncertainty from time $0$ to time $T$.
Uncertainty is modeled by a complete probability space $(\Omega, \F,\P)$.
Each $\omega \in \Omega$ is a state of nature which is a complete description of one possible realization of all exogenous sources of uncertainty from time $0$ to time $T$.
The sigma-field $\F$ is the collection of events which are distinguishable at time $T$ and $\P$ is a probability measure on $(\Omega,\F)$.
The probability space $(\Omega,\F,\P)$ is endowed with a filtration $\mathbb{F}=\{ \F(t)\colon t\in [0,T]\}$ which represents the time evolution of the agents' knowledge about the states of nature.
We assume that $\F(0)$ is $\P$-almost surely trivial and that $\mathbb{F}$ satisfies the usual conditions of right-continuity and completeness.
A process is said optional if it is $\mathcal{O}$-measurable where $\mathcal{O}$ is the sigma-field on $\Omega \times [0,T]$ generated by right-continuous $\mathbb{F}$-adapted processes with left-limits.

\subsection{Consumption space}

There is a single consumption good available for consumption at any time $t\in [0,T]$.
The set of positive, nondecreasing and right-continuous functions from $[0,T]$ to $\R_+$ is denoted by $M_+$.
We represent the consumption bundle of an agent by a process $x :(\omega,t) \mapsto x(\omega,t)$, where $x(\omega,t)$ (sometimes denoted by $x_t(\omega)$) represents the cumulative consumption from time $0$ to time $T$ and satisfies
\begin{description}
\item[(a)] for each $\omega \in \Omega$, the function $x(\omega)$ belongs to $M_+$,
\item[(b)] for each $t\in [0,T]$, the random variable $x_t$ is $\F(t)$-measurable and $x_T$ belongs to $L^p(\P)$
\end{description}
where $1\leq p< +\infty$.

The set of ($\P$-equivalent classes of) mappings $x : \Omega \rightarrow M_+$ such that the process $(\omega,t) \mapsto x(\omega,t)$ satisfies (a) and~(b) is denoted by $E_+$ and the linear span of $E_+$ will be denoted by $E$.
The space $E_+$ is called the consumption space and $E$ is called the commodity space.
Observe that any consumption bundle $x$ in $E_+$ is an $\mathbb{F}$-adapted process having right-continuous and bounded variation sample paths and therefore can be assimilated with an optional random measure denoted by $dx$.
If $z$ belongs to $E$ then there exist $x,y$ in $E_+$ such that $z=x-y$.
We can endow $E$ with the linear order $\geq$ defined by the cone $E_+$ in the sense that $y\geq x$ if $y-x$ belongs to $E_+$.
If $y$ belongs to $E_+$ then the order interval $[0,y]$ is defined by $[0,y]:=\{ x\in E \colon x\in E_+ \et y-x \in E_+ \}$.
The space $E$ endowed with the partial order defined by $E_+$ is a linear vector lattice (see \cite[Proposition~1]{Martinsda}).

\begin{remark}
Observe that if $x,y$ are vectors in $E$ such that $y \geq x$ then there exists $\Omega^* \in \F$ with $\P\Omega^*=1$ and such that for each $\omega \in \Omega^*$, the function $t \longmapsto y(\omega,t)-x(\omega,t)$ is nondecreasing with $y(\omega,0) -x(\omega,0) \geq 0$.
In particular we have for each $\omega \in \Omega^*$,
\[
y(\omega,t) \geq x(\omega,t) \quad \forall t\in [0,T].
\]
\end{remark}

\subsection{Topologies}

Since $0\leq x_t \leq x_T$ and $x_T \in L^1(\P)$ for every $x \in E_+$, the space $E$ is a subspace of $L^1(\mathcal{O},\P\otimes \kappa)$ where $\kappa = \lambda + \delta_T$ with $\lambda$ the Lebesgue measure on $\B$ the Borelian sigma-algebra on $[0,T]$ and $\delta_T$ the Dirac measure on $T$.
Following \cite{HindyHuang92} we consider on $E$ the restriction of the $L^p(\mathcal{O},P\otimes \kappa)$-norm, i.e., we consider the norm $\norm{\cdot}$ defined by
\[
\forall x\in E, \quad \norm{x}=\left[ \E \int_{[0,T]} |x(t)|^p \kappa(dt) \right] ^{\frac{1}{p}} =\left[ \E \int_{[0,T]} |x(t)|^p dt + \E|x(T)|^p \right] ^{\frac{1}{p}}.
\]
It is argued in~\cite{HindyHuang92} that this norm, called intertemporal norm, induces a topology on the set of consumption bundles that exhibits intuitive economic properties, in particular it captures the notion that consumption at adjacent dates are almost perfect substitutes except possibly at information surprises.
Usually, we refer to the topology generated by the intertemporal norm when we speak about continuity, open sets, etc.
Occasionally, we will use other topologies as well, though.
For $z \in E$ and fixed $\omega \in \Omega$, the function $z(\omega)$ can be assimilated with a signed measure $d[z(\omega)]$ on the time interval $[0,T]$.
We denote by $\normtv{z(\omega)}$ the total variation of the measure $d[z(\omega)]$ (and we will drop the $\omega$ frequently, as usual).
The expectation of the total variation of $z$ leads to the strong topology on $E$ as given by the norm
\[
\normstrong{z}:=\E \normtv{z}.
\]
Note that convergence in the strong topology entails convergence in the intertemporal topology.
Moreover, $E$ is a topological vector lattice when endowed with the order generated by $E_+$ and the strong topology.

If $h$ is a random variable and $\tau$ a stopping time in $\mathcal{T}$, we denote by $\delta_\tau h$ the simple random measure that delivers $h(\omega)$ units of the consumption good at time $\tau(\omega)$ and nothing elsewhere.
In particular $\delta_\tau$ is the Dirac measure on $\tau$.

\subsection{Prices}

The weakest notion of a price is that of a nonnegative linear functional on $E$.
The algebraic dual (the space of linear functional from $E$ to $\R$) is denoted by $E^\star$ and $E^\star_+$ denotes the cone of nonnegative linear functionals, i.e., $\pi \in E^\star$ is nonnegative if $\pi(x) \geq 0$ for every $x\in E_+$.
If $B(T)$ denotes the space of bounded functions defined on $[0,T]$ then we let $L^q(\P,B(T))$ denote the space (up to $\P$-indistinguishability) of all $\F\otimes\B$-measurable processes $\psi: \Omega \times [0,T] \rightarrow \R$ such that the function
\[
\omega \rightarrow \sup_{t\in [0,T]} |\psi(\omega,t)|
\]
belongs to $L^q(\P)$ where $q\in (1,+\infty]$ is the conjugate of $p$.
There is a natural duality $\ps{\cdot}{\cdot}$ on $L^p(\P,B(T)) \times E$ defined by
\[
\ps{\psi}{z} = \E \int_{[0,T]} \psi(t) dz(t).
\]
The space of processes in $L^q(\P,B(T))$ that are optional is denoted by $F$ and we denote by $F_+$ the order dual cone, i.e., \[
F_+ := \{ \psi \in F \ \colon \ \ps{\psi}{x}\geq 0, \quad \forall x\in E_+ \}.
\]
The pair $\ps{F}{E}$ is a Riesz dual pair (see \cite[Proposition~1]{Martinsda}) and a process $\psi \in F$ belongs to $F_+$ if and only if $\psi(t) \geq 0$ for every $t\in [0,T]$.
To each nonnegative process $\psi \in F_+$ we can consider the nonnegative linear functional $\ps{\psi}{\cdot}$ in $E^\star_+$ defined by
\[
\forall z\in E, \quad \ps{\psi}{z} = \E \int \psi dz.
\]
By abuse of notations, we still denote $F_+$ (and $F$) the space of linear functionals associated to processes in $F_+$ (resp. $F$).
If a price $\pi \in E^\star_+$ is represented by an optional process $\psi \in F_+$ in the sense that $\pi=\ps{\psi}{\cdot}$, then the process $\psi$ is called a state price.
In that case the duality product $\ps{\psi}{x}$ is the value of the consumption bundle $x\in E_+$ under the price $\psi$ where $\psi(\omega,t)$ is interpreted to be the time $0$ price of one unit of consumption at time $t$ in state $\omega$, per unit of probability.

\section{Compatible Prices}\label{section:compatible}

In general, prices in $F_+$ will not be compatible  with the notion of intertemporal substitution as they might assign very different prices to consumption plans that are close in the intertemporal topology.
One might therefore aim to find prices in the topological dual $(E,\norm{\cdot})'$ of $E$.
As shown by \cite{HindyHuang92}, every linear functional $\pi \in (E,\norm{\cdot})'$ continuous for the intertemporal norm can be represented\footnote{In the sense that $\pi=\ps{\psi}{\cdot}$.} by a semimartingale $\psi$ satisfying
\[
\psi_t = A_t + M_t
\]
where $A$ is an adapted process with absolutely continuous sample path satisfying
\[
A' \in L^q(\mathcal{O},\P \otimes \kappa) \et A'_T \in L^q(\P)
\]
and $M$ is the martingale defined by
\[
M_t = \E[ -A'_T - A_T \vert \F_t], \quad \forall t\in [0,T].
\]
We denote by $K$ the space of processes $\psi$ representing linear functionals in $(E,\norm{\cdot})'$.\footnote{Observe that $K$ is a subset of $F$.}
However, there is in general no hope to obtain equilibrium prices in $K$ as it is a not a lattice.
On the other hand, all that counts for equilibrium theory are linear functionals restricted to the \emph{consumption set} $E_+$, the positive cone of the commodity space.
So we relax the requirement of continuity on the whole space and aim only for continuity on the consumption set $E_+$.
A price is called \emph{compatible} if if is continuous with respect to the intertemporal topology on $E_+$.
Denote the set of compatible prices by $H_+$.

This leads to two questions:
     \begin{itemize}
     \item What is the structure of compatible prices?
     \item Under which conditions do equilibria with compatible prices exist?
     \end{itemize}
Let us answer the first question.

\begin{theorem}\label{ThmCharCompPrices}
A nonnegative linear functional $\pi \in E^\star_+$ is a compatible price if and only if it can be written as
\[
\pi(x)=\E \int \psi dx, \quad \forall x \in E_+
\]
for a nonnegative, rightcontinuous processes $\psi$ with left limits that satisfies the following conditions:
  \begin{itemize}
    \item The process $\psi$ is the optional  projection\footnote{That is $\psi_\tau = \E[\xi_\tau \vert \F_\tau]$ for every stopping time $\tau \leq T$.} of a (not necessarily adapted) continuous process $\xi$ with
    \[
    \E \sup_{t\in[0,T]} |\xi_t| < \infty.
    \]
    \item The process
    \[
    \psi^*=\sup_{t\in[0,T]} \psi_t
    \]
    belongs to $L^q(\F, \P)$.
\end{itemize}
\end{theorem}

We denote by $H$ the space the rightcontinuous processes with left limits that are bounded in $L^q$ and are the optional projection of a continuous process bounded in $L^1$.
Observe that the space $H$ is a subspace of $F$ containing $K$.
If we let $H_+=H\cap F_+$ then a nonnegative linear functional $\pi$ is compatible if only if it can be represented by a process $\psi$ in $H_+$, i.e., $\pi=\ps{\psi}{\cdot}$.
We give the proof of Theorem~\ref{ThmCharCompPrices} for the case $p=1$.
Later, we indicate how to obtain the result for $p>1$.

\subsection{Sufficiency}

First, we show that every element that satisfies the conditions of the theorem  induces a compatible price.

\begin{lemma}
Let $\psi$ be a cadlag process that satisfies the assumptions in Theorem~\ref{ThmCharCompPrices}.
Then the mapping $\pi=\ps{\psi}{\cdot}$ is a compatible price.
\end{lemma}

\begin{proof}
Let $M \geq 0$ be an upper bound for $\psi$, i.e., $M \geq \sup_t \psi_t $ almost surely.
As processes $z \in E$ have integrable variation, $\ps{\psi}{\cdot}$ is well defined on $E$:
\[
\forall z\in E, \quad \left|\E \int \psi dz \right| \leq  M \normstrong{z} < \infty.
\]
Since $\ps{\psi}{\cdot}$ is obviously linear and nonnegative, it belongs to $E^\star_+$.

Let $\xi$ be a continuous process which optional projection $^o\xi$ coincides with $\psi$.
We first establish continuity of the functional $\ps{\psi}{\cdot}$ on the space
\[
E^k_+:=\left\{x \in E_+ : x_T \leq k \quad \as \right\}
\]
for arbitrary $k>0$.
Let  $(x^n) \subset E^k_+$ be a sequence converging to some $x \in E^k_+$ for the intertemporal norm.
As we have
\[
\E \int \psi dz = \E \int \xi dz
\]
for all $z \in E$, it is enough to prove that
\[
\lim_n \E \int \xi dx^n = \E \int \xi dx.
\]
Suppose this is not true.
Then there is a subsequence $(y^n)$ of $(x^n)$ such that
\[
\lim_{n} d^n:= \E \int \xi dy^n = d \not= c :=\E \int \xi dx.
\]
Without loss of generality, we can assume that on a set of probability $1$, the sequence $(y^n)$ converges weakly in the sense of measures on the time axis to $x$ (see Lemma~1 in ~\cite{Martinsda} or \cite[Proposition~5]{HindyHuangKreps92}).
Then we have $\lim_n \int \xi dy^n =  \int \xi dx$ almost surely because $\xi$ is continuous.
From
\[
\left|\int \xi dy_n\right| \leq k \sup_{t \in[0,T]} |\xi_t|
\]
and the assumption on $\xi$, we get by dominated convergence that $\lim_n d^n =c$: a contradiction.

Now let $(x^n)$ be an arbitrary sequence in $E_+$ that converges to $x$.
For each $k\in \N$, we let $x^n_k$ and $x_k$ the optional random measures defined by
\[
dx^n_k=dx^n\wedge [\delta_0 k] \et dx_k = dx \wedge [\delta_0 k].
\]
Observe that for every $t\in [0,T]$ we have $x^n_k(t)=\min\{x^n(t),k\}$ and $x_k(t)=\min\{C(t),k\}$.
It follows immediately from dominated convergence that
\begin{equation}\label{Eqn2}
\lim_{k} \normstrong{x_k-x} = 0
\end{equation}
and
\begin{equation}\label{Eqn3}
\forall k\in \mathbb N, \quad\lim_{n} \norm{x^{n}_k-x_k} = 0.
\end{equation}
Observe that for every $(k,n)$ we have
\begin{equation}\label{eq:nk-n}
\left\vert \ps{\psi}{x^{n}_k - x^n} \right \vert \leq M \normstrong{x^{n}_k - x^n} = M \E[x^n - k]^+ \leq M \norm{x^n - x} + M \E[x_k - k]^+.
\end{equation}
For $\epsilon>0$, relation \rref{Eqn2} allow us to find $k_0$ such that
\begin{equation}\label{eq:c-ck}
\left| \ps{\psi}{x - x_k} \right| \leq M \normstrong{x - x_k} < \epsilon \et M \E[x_k - k]^+ \leq \varepsilon.
\end{equation}
Now fix $k=k_0$, it follows from \rref{eq:nk-n} and \rref{eq:c-ck} that for all $n \in \N$
\begin{eqnarray}
  \left| \ps{\psi}{x-x^n} \right| &\leq & \left| \ps{\psi}{x-x_k}\right| + \left| \ps{\psi}{x_k-x^{n}_k} \right| +  \left| \ps{\psi}{x^{n}_k-x^{n}} \right|\\
 &\leq &  2 \epsilon +\left| \ps{\psi}{x_{k}-x^{n}_k}\right| + M \norm{x^n - x}.
\end{eqnarray}
By the fact that $\pi$ is continuous on $E^k_+$ and \rref{Eqn3}, we can choose $n_0$ such that for all $n \geq n_0$
\[
\left| \ps{\psi}{x_{k}-x^{n}_k} \right| < \varepsilon \et M \norm{x^n - x} \leq \varepsilon
\]
and we finally obtain
\[
\left| \ps{\psi}{x-x^n} \right|  < 4 \epsilon
\]
for $n \geq n_0$.
This shows that $\pi$ is continuous on $E_+$.
\end{proof}

\subsection{Necessity}

The converse is the much more demanding part.
Given a compatible price $\pi \in H_+$, we have to find a \emph{density} $\psi $ that represents $\pi$.
We will frequently use the following continuity lemma that yields suitable upper bounds.

\begin{lemma}\label{LemStrongCont}
Compatible prices are continuous with respect to the strong topology.
\end{lemma}

\begin{proof}
As the strong topology is stronger than the intertemporal topology, a compatible price $\pi$ is $\normstrong{\cdot}$--continuous on $E_+$.
But the space $E$ is a topological vector lattice with respect to the strong topology.
Hence, the lattice operations are continuous with respect to the strong topology.
It follows that $\pi$ is $\normstrong{\cdot}$--continuous on the whole space $E$.
\end{proof}

As the space $\left(E, \normstrong{\cdot}\right)$ is a Banach space, the preceding lemma yields a constant $K>0$ such that
\begin{equation}\label{EqnBoundedPi}
\forall z\in E, \quad \left| \pi(z) \right| \leq K \normstrong{z}.
\end{equation}

Denote by $\mathcal{T}$ the set of all stopping times $\tau \leq T$.
A $\mathcal{T}$--system is a family $(z^\tau)_{\tau \in \mathcal{T}}$ of random variables that
satisfy (see \cite{DellacherieLenglart80})
\begin{enumerate}
  \item consistency: for $\sigma,\tau$ stopping times  $z^\tau=z^\sigma$ on the set $\{\tau=\sigma\}$,
  \item measurability: every random variable $z^\tau$ is $\cal F_\tau$--measurable.
\end{enumerate}
Fix a random variable $\tau \in \mathcal{T}$.
Define a linear mapping $Q^\tau$ from $L^1(\F_\tau,\P)$ into $\mathbb R$ by setting
\[
\forall Z \in L^1(\F_\tau,\P), \quad Q^\tau(Z)=\pi\left(\delta_\tau Z\right).
\]
Being a continuous linear mapping, it can be represented by a random variable $z^\tau \in L^\infty(\F_\tau,\P)$ such that
\[
\forall Z \in L^1(\F_\tau,\P), \quad Q^\tau(Z)=\E(Z  z^\tau).
\]
As $\pi$ is nonnegative we actually have $z^\tau \geq 0$ a.s.

\begin{claim}
The family $(z^\tau)_{\tau \in \mathcal{T}}$ forms a $\mathcal{T}$--system.
\end{claim}

\begin{proof}
It is sufficient to show that $z^\sigma \1_{\{\sigma=\tau\}}=z^\tau \1_{\{\sigma=\tau\}}$ almost surely.
As both $z^\sigma \1_{\{\sigma=\tau\}}$ and $z^\tau \1_{\{\sigma=\tau\}}$ are $\F_{\sigma \wedge \tau}$--measurable,  it is enough to show
\[
\forall Z \in L^1(\F_{\sigma \wedge \tau},\P),\quad \E z^\sigma \1_{\{\sigma=\tau\}} Z = \E z^\tau \1_{\{\sigma=\tau\}} Z.
\]
Take such a $Z$ in $L^1(\F_{\sigma \wedge \tau},\P)$.
Then
\begin{align*}
\E  z^\sigma  \1_{\{\sigma=\tau\}} Z
&= Q^\sigma\left( \1_{\{\sigma=\tau\} } Z\right)\\
&= \pi\left(\delta_\sigma \1_{\{\sigma=\tau\} } Z\right)\\
&= \pi\left(\delta_\tau \1_{\{\sigma=\tau\} }Z\right)\\
&= Q^\tau\left( \1_{\{\sigma=\tau\} } Z\right)\\
&= \E  z^\tau  \1_{\{\sigma=\tau\}} Z\,.
\end{align*}
This concludes the proof.
\end{proof}

The question is: can we find a process $(\psi_t)_{t\in [0,T]}$ such that $\psi_\tau = z^\tau$ almost surely for all stopping times $\tau \in \mathcal{T}$?
Such a question is called a \emph{problem of aggregation} in the th\'eorie g\'en\'erale of stochastic processes.
In general, aggregation is not possible without some continuity requirement (see \cite{DellacherieLenglart80}).
Therefore, we establish the following lemma.

\begin{claim}\label{LemContExp}
The $\mathcal{T}$--system $(z^\tau)_{\tau \in \mathcal{T}}$ is continuous in expectation in the sense that
\[
\lim_n \E z^{\tau_n}= \E z^\tau
\]
for all sequences of stopping times $\left(\tau_n\right)$ with $\lim_n \tau_n = \tau$.
\end{claim}

\begin{proof}
Let $(\tau_n)$ be a sequence of stopping times satisfying $\lim \tau_n = \tau$.
Then, the sequence of optional random measures $(\delta_{\tau_n})$ converge to $\delta_\tau$ in the intertemporal topology.
Continuity of the price functional $\pi$ implies
\begin{equation*}
\lim_n  \E z^{\tau_n} = \lim_n \pi\left(\delta_{\tau_n}\right) = \pi\left(\delta_{\tau}\right)= \E z^{\tau}.
\end{equation*}
\end{proof}

\begin{claim}\label{claim:uniformly integrable}
There exists a nonnegative, adapted and cadlag process $\psi \in L^\infty(\P,\B)$ that aggregates (or recollects) $(z^\tau)_{\tau \in \mathcal{T}}$ in the sense that $\psi_\tau=z^\tau$ almost surely for every stopping time $\tau \in \mathcal{T}$.
\end{claim}

\begin{proof}
By \cite[Theorem~6]{DellacherieLenglart80}, every nonnegative $\mathcal{T}$--system can be aggregated by an optional process $\psi$.
The process $\psi$ is nonnegative because so are every $z^\tau$.
From Claim~\ref{LemContExp} the process $\psi$ is continuous in expectation.
If we prove that $\psi$ is uniformly integrable, then we can apply \cite[Theorem 48]{DM78}\footnote{The theorem is formulated for bounded processes only. The comment~50(f) in \cite{DM78} shows that it is enough to have uniform integrability.} to conclude that  $\psi$ is cadlag.
We first prove that $\psi$ is bounded in $L^1$.
Let $\tau$ be a stopping time and observe that
\[
0 \leq \E \psi_\tau = \E z^\tau = \pi(\delta_\tau).
\]
The process $\delta_\tau$ belongs to $E$ and \rref{EqnBoundedPi} yields $\pi(\delta_\tau) \leq K$ implying that $\E \psi_\tau < \infty$.
To establish uniform integrability, we have to show that for all $\varepsilon >0$ there exists $\delta>0$ such that for all sets $A \in \F$ with $\P(A) \leq \delta$ one has $\E\1_{A} \psi_\tau \leq \varepsilon$ for every stopping time $\tau$.
For $A \in \F$ and a stopping time $\tau$, let $c=\delta_\tau \E[\1_A \vert \F_\tau]$.
Since the process $c$ belongs to $E$ we have
\[
\E \1_A \psi_\tau = \E ( \E[\1_A\vert \F_\tau] \psi_\tau) = \pi(c) \leq K \normtv{c} = K \P(A).
\]
Setting $\delta=\varepsilon/K$, we obtain uniform integrability.
\end{proof}

\begin{claim}\label{claim:bounded}
The process $\psi$ is bounded.
\end{claim}

\begin{proof}
Fix $\alpha>0$ and let $\tau$ be the stopping time
\[
\tau =
\left\{
\begin{array}{ll}
\inf\left\{t \ge 0 : \psi_t \ge n\right\} & \textrm{if} \ \sup_{0 \le t \le T} \, \psi_t \geq \alpha \\
~\\
\infty & \textrm{elsewhere}
\end{array}
\right.
\]
Recall that the random variable $\psi_\tau$ is given  by $\psi_\tau(\omega) = \psi_{\tau(\omega)}(\omega) \1_{\{ \tau< \infty\}}$.
Consider the optional random measure $c=\delta_{\tau} \1_{\{ \tau < \infty\}}$. We have
\[
\pi(c)=\E z^{\tau} \1_{\{ \tau < \infty\}} =\E \psi_{\tau} \1_{\{\tau < \infty \}}.
\]
Since $\psi$ is cadlag, we get $\pi(c) = \E \psi_{\tau} \1_{\{\tau < \infty \}} \geq \alpha \P \{\tau < \infty\}.$
On the other hand, \rref{EqnBoundedPi} yields
\[
\pi(c) \leq K \normstrong{c} = K \E \normtv{\delta_{\tau} \1_{\{ \tau < \infty\}} } = K  \P \{ \tau < \infty \}.
\]
Choosing $\alpha >K$ then shows that $\P \{\tau < \infty\} =0$.
Hence the process $\psi$ is bounded, i.e., $\psi \in L^\infty(\P,B(T))$.
\end{proof}

In general, $\psi$ is not going to be continuous. However, we have the following result that goes back to \cite{Bismut78} and \cite{Emery78}.

\begin{claim}
There exists a \emph{not necessarily adapted} continuous process $\xi $ with
\[
\E \sup_{t \in [0,T]} | \xi_t | < \infty
\]
whose optional projection is $\psi$, that is
\[
\psi_\tau = \condexp[\F_\tau]{\xi_\tau}
\]
for all stopping times $\tau \in \mathcal{T}$.
\end{claim}

\begin{proof}
This is the main theorem in \cite{Bismut78} and \cite{Emery78}.
According to the notations in \cite{Bismut78} and \cite{Emery78}, we have to check the conditions that $\psi$ is regular and of class (D).
As $\psi$ is bounded, it is of class (D).
A process is regular if and only if the predictable projection of $\psi$ is equal to $\psi_-$.
This is equivalent to continuity in expectation from below (see \cite[50(d)]{DM78}).
As $\psi$ is even continuous in expectation, it is regular.
\end{proof}

\begin{claim}\label{LemRepLinfty}
For every bounded consumption plan $x\in E_+$ with $x_T \in L^\infty(\P)$ we have $\pi(x) = \ps{\psi}{x}$, i.e., \[
\pi(x) = \E \int \psi dx.
\]
\end{claim}

\begin{proof}
By construction, we have for every stopping time $\tau$  and $\F_\tau$--measurable random variable~$h$
\[
\pi(\delta_\tau h)=\E z^\tau h = \E \psi_\tau h = \E \xi_\tau h.
\]
Via linearity, we obtain $\pi(z)=\ps{\psi}{z}$ for every simple random measure $z$.
As simple random  measures are dense with respect to the intertemporal norm in $E_+$ and $\xi$ is continuous, we get the result for optional random measures with bounded variation in $L^\infty(\P)$ (observe that \cite{Bismut78} obtains this in his proof).
\end{proof}

Since $\psi$ belongs to $L^\infty(\P,B(T))$, the random variable $\psi^\star$ defined by
\[
\psi^\star = \sup_{t\in [0,T]} \psi_t
\]
belongs to $L^\infty(\P)$.
It follows that for every consumption plan $x\in E_+$ the quantity $\ps{\psi}{x}$ is well defined as
\[
\ps{\psi}{x} = \E \int \psi dx \leq \E \psi^* x_T < \infty.
\]
We can now prove that $\pi(x)=\ps{\psi}{x}$ for every $x \in E_+$.
From Claim \ref{LemRepLinfty}, we know that $\pi(x)=\ps{\psi}{x}$ for all bounded $x$ in $L^\infty(\P,B(T))$.
Now let $x \in E_+$ be given.
Set $dx_n=dx \wedge n\delta_0$, i.e., $x_n(t) = \min\{x_t,n\}$ for every $t\in [0,T]$ .
For each $n$ the optional random measure $x_n$ is bounded and the sequence $(x_n(\omega))$ converges for the total variation norm $\normtv{.}$ to $x(\omega)$ from below for all $\omega$.
Consequently, for all nonnegative measurable functions $f$ we have
\[
\lim_n \int f dx_n =\int f dx.
\]
In particular, we have $\lim_n \int \psi dx_n = \int \psi dx$ almost surely.
By monotone convergence, we obtain
\[
\lim_n \pi(x_n)=\lim_n \E\int \psi dx_n = \E\int \psi dx =\ps{\psi}{x}
\]
and as $\pi$ is continuous with respect to the strong topology, $\pi(x)=\ps{\psi}{x}$ follows.
This concludes the proof of the theorem.

\subsection{The proof for $p>1$}

For $p>1$, the proof follows almost verbatim the above proof for $p=1$.
However, one cannot use the argument given above that establishes boundedness of the process $\psi$.
Instead, one has to use a different argument to prove that the supremum of $\psi$ is in $L^q$.
This argument is given next.

\begin{claim}
The  supremum
\[
\psi^*:=\sup_{t\in [0,T]} \psi_t
\]
satisfies
\[
\forall H\in L^p(\P), \quad \E \psi^* |H| \leq (K+1) \norm{H}_{L^p}.
\]
In particular the random variable $\psi^*$ belongs to $L^q$.
\end{claim}

\begin{proof}
Let $S$ be a random time (not necessarily a stopping time) and $h \in L^p_+(\P)$.
Denote by $z=\delta_S h$ and by $x=\left(z\right)^o$ its optional dual projection.
Then $x$ is an optional random measure and
\[
\E \psi_S h  = \E \int \psi dz = \E \int \psi dx = \pi(x)\leq K \normstrong{x}.
\]
The process $x$ is nondecreasing and $\F=\F_T$, hence
\[
\normstrong{x} = \E \normtv{x} = \E x_T = \E h \leq \norm{h}_{L^p}.
\]
Let $S$ be a cross--section of the set
\[
\left\{(\omega,t): \psi_t(\omega)\geq \psi^\star(\omega) - 1 \right\}.
\]
Then we have
\[
\E \psi^* h  \leq \E (\psi_S + 1) h \leq (K+1) \norm{h}_{L^p}.
\]
\end{proof}

\section{Equilibria with Compatible Prices}\label{section:equilibrium}

Each agent $i$ is characterized by a utility function $V^i : E_+ \longrightarrow \R$ which represents his preference relation on the space $E_+$ of consumption patterns and by a vector $e^i \in E_+$ which represents the cumulative income stream (initial endowment).
An economy is a pair
\[
\mathcal{E}=(\bm{V},\bm{e})
\]
where $\bm{V}=(V^i)_{i\in I}$ and $\bm{e}=(e^i)_{i\in I}$.
We let $e=\sum_{i\in I} e^i$ denote the aggregate endowment and if $x \in E_+$ the set $\{ y\in E_+ \ \colon \ V^i(y) > V^i(x) \}$ is denoted by $P^i(x)$.
An allocation is a vector $\bm{x}=(x^i)_{i\in I}$ where $x^i\in E_+$.
It is said feasible or attainable if $\sum_{i\in I} x^i = e$.
The set of attainable allocations is denoted by~$\mathcal{A}$.

\subsection{Equilibrium concepts}

We define hereafter the standard notion of Arrow--Debreu equilibrium.

\begin{definition}
The pair $(\psi,\bm{x})$ of a price process $\psi$ and an allocation $\bm{x}$ is called an Arrow--Debreu equilibrium if
\begin{description}
\item[(a)] the price process $\psi$ belongs to $F_+$ and $\ps{\psi}{e} >0$;
\item[(b)] the allocation $\bm{x}$ is attainable, i.e., $\bm{x}\in \mathcal{A}$; and
\item[(c)] for each agent $i$, the consumption plan $x^i$ maximizes agent i's utility over all consumption plans $y$ satisfying the budget constraint $\ps{\psi}{y} \leq \ps{\psi}{e^i}$, i.e., \[
x^i \in \mathrm{argmax} \{ V^i(y) \ \colon \ y \in E_+ \quad \textrm{and} \quad \ps{\psi}{y} \leq \ps{\psi}{e^i} \}.
\]
\end{description}
\end{definition}

A possible interpretation is that a complete set of markets open at the initial date $t=0$ for consumption good delivery at any date in any state of nature.
Markets are assumed to be competitive in the sense that agents take the price functional $\ps{\psi}{\cdot}$ as given.
Each agent can sell his initial endowment $e^i$ and buy a consumption plan $x \in E_+$ as far as he can afford it, i.e., $\ps{\psi}{x} \leq \ps{\psi}{e^i}$.
The real number $\ps{\psi}{x}$ is interpreted as the price at time $t=0$ of the consumption claim $x$, and therefore the real number $\psi(\omega,t)$ is interpreted as the time $t=0$ price (per unit of probability) of the contract that promises to deliver one unit of the unique good at time $t$ in state $\omega$.

\begin{remark}
Observe that if $(\psi,\bm{x})$ is an equilibrium then the budget constraints are binding, i.e., for each $i$, we have $\ps{\psi}{x^i}=\ps{\psi}{e^i}$.
\end{remark}

As usual in general equilibrium literature, we consider the following list of \emph{standard} assumptions.
\begin{assumption}[C]\label{hyp:C}
For each agent $i$,
\begin{description}
\item[(C.1)] the initial endowment $e^i$ belongs to $E_+$ and is not zero, i.e., $e^i>0$,
\item[(C.2)] the utility function $V^i$ is concave,
\item[(C.3)] the utility function $V^i$ is norm continuous.\footnote{Actually, it is sufficient to assume that $V^i$ is upper semi-continuous on the order interval $[0,e]$. That is, if $(x_n)_{n\in \N}$ is a sequence in $[0,e]$ which norm-converges to $x$ in $[0,e]$, then
\[
\limsup_{n\rightarrow \infty} V^i(x_n) \leq V^i(x).
\]}
\end{description}
\end{assumption}

We recall a well-known property of optimality for allocations.

\begin{definition}\label{def:core}
An attainable allocation $\bm{x} \in \mathcal{A}$ is said to be an {\it Edgeworth} equilibrium if there is no $0\neq \bm{\lambda} \in (\Q\cap [0,1])^I$ and some allocation $\bm{y}$ such that $V^i(y^i) > V^i(x^i)$ for each $i$ with $\lambda^i>0$ and satisfying $\sum_{i\in I} \lambda^i y^i = \sum_{i\in I} \lambda^i e^i$.
\end{definition}

The reader should observe that this concept is ``price free" in the sense that it is an intrinsic property of the commodity space.
It is proved in \cite{Martinsda} that every economy satisfying Assumption~C admits an Edgeworth equilibrium.
It is straightforward to check that every Arrow--Debreu equilibrium is an Edgeworth equilibrium.
The main difficulty consists in proving the converse.

\subsection{Properness of preferences}

We propose to follow the classical literature\footnote{We refer, among others, to \cite{MAS}, \cite{RZ}, \cite{YZ}, \cite{ABB87a}, \cite{ABB87b}, \cite{Zame87}, \cite{RIC}, \cite{AM}, \cite{MR},  \cite{MasColellZame91}, \cite{POD}, \cite{AndersonZame97}, \cite{AndersonZame98}, \cite{TOU98}, \cite{DF}, \cite{TOU99}, \cite{ATY01}, \cite{ShannonZame}, \cite{FLO03}, \cite{AliprantisMonteiroTourky}, \cite{AFT04} and \cite{AFT05}.} dealing with infinite dimensional commodity-price spaces by introducing the concept of proper economies.
It is a well-known fact that without some properness hypotheses on preferences, equilibrium existence may fail when the positive cone of the commodity space has empty interior.

\begin{definition}[$\tau$-properness]\label{def:tau-proper}
Let $\tau$ be a Hausdorff locally convex linear topology on $E$.
An economy $(\bm{V},\bm{e})$ is $\tau$-proper if for every Edgeworth equilibrium $\bm{x}$, for each $i$, there is a set $\widehat{P}^i(x^i)$ such that
\begin{description}
\item[(i)] the vector $x^i+e$ is a $\tau$-interior point of $\widehat{P}^i(x^i)$,
\item[(ii)] the set $\widehat{P}^i(x^i)$ is convex and satisfies the following additional convexity property
\[
\forall z\in \widehat{P}^i(x^i) \cap E_+, \quad \forall t\in (0,1), \quad tz +(1-t)x^i \in \widehat{P}^i(x^i) \cap E_+
\]
\item[(iii)] we can extend preferences in the following way
\[
\widehat{P}^i(x^i)  \cap E_+ \cap A_{x^i} \subset P^i(x^i) \subset \widehat{P}^i(x^i) \cap E_+
\]
where $A_{x^i} \subset E$ is a radial set at $x^i$.\footnote{A subset $A$ of $E$ is radial at $x\in A$ if for each $y\in E$, there exists $\bar{\alpha} \in (0,1]$ such that $(1-\alpha) x + \alpha y$ belongs to $A$ for every $\alpha \in [0,\bar{\alpha}]$.}
\end{description}
\end{definition}
We say that an economy is {\it strongly} $\tau$-proper if condition (iii) in Definition~\ref{def:tau-proper} is replaced by the following condition~(iii'):
\begin{description}
\item[(iii')] we can extend preferences in the following way
\[
\widehat{P}^i(x^i)  \cap E_+ = P^i(x^i).
\]
\end{description}
Strong $\tau$-properness was introduced by \cite{TOU99} and is used, among others, by \cite{ATY01}, \cite{AFT04} and \cite{AFT05}.
We refer to \cite{ATY} for a comparison of the different notions of properness used in the literature.
Observe that if $\mathcal{E}=(\bm{V},\bm{e})$ is an economy (satisfying the following monotonicity Assumption~M) such that for each $i$, it is possible to extend $V^i$ to a $\tau$-continuous and concave function $\widehat{V}^i : E \longrightarrow \R$, then the economy is $\tau$-proper.\footnote{Take $\widehat{P}^i(x) := \{ y\in E \ \colon \ \widehat{V}^i(y) > \widehat{V}^i(x) \}$.}
In other words, $\tau$-properness can be seen as a strengthening of $\tau$-continuity.
Moreover, $\tau$-properness is slightly weaker than strong $\tau$-properness.
However this slight difference is crucial in order to compare properness with the existence of smooth sub-gradients.\footnote{See also Assumption~A.7 in \cite{ShannonZame}.}
We borrow the following definition of smooth sub-gradients from \cite{BankRiedel02} (see also \cite{Martinsda}).
Recall that $K$ is the space of processes in $F$ that represent linear functionals on $E$ that are norm continuous.

\begin{definition}\label{def:U}
An economy $(\bm{V},\bm{e})$ has smooth sub-gradients in $K$ if for each $i$, for every $x \in E_+$, there exists a nonnegative optional process $\nabla V^i(x) \in K_{+}=K\cap F_+$ with
\begin{description}
\item[(U.1)] for each $j\in I$, we have $\ps{\nabla V^i(x)}{e^j} >0$,
\item[(U.2)] the vector $\nabla V^i(x)$ satisfies the subgradient property
\[
\forall y \in E_+, \quad V^i(y) - V^i(x) \leq \ps{\nabla V^i(x)}{y- x}
\]
\item[(U.3)] this subgradient is continuous in the sense that,
\[
\forall y \in E_+, \quad  \lim_{\varepsilon \downarrow 0} \ps{\nabla V^i(\varepsilon y + (1-\varepsilon) x)}{y -x} = \ps{\nabla V^i(x)}{y -x}.
\]
\end{description}
\end{definition}

\begin{remark}\label{rem:increasing}
Let $\mathcal{E}=(\bm{V},\bm{e})$ be an economy.
Preferences of agent $i$ are said increasing if $V^i(x + y) \geq V^i(x)$ for every $x$, $y$ in $E_+$; strictly increasing if $V^i(x + y) > V^i(x)$ for every $x$, $y$ in $E_+$ with $y\neq 0$.
Note that if $\mathcal{E}$ satisfies Assumption~U, then preferences of agent $i$ are increasing; they are strictly increasing if and only if $\nabla V^i(x)$ is strictly positive  for every $x \in E_+$.
\end{remark}

\begin{remark}\label{rem:desirable}
Let $(\bm{V},\bm{e})$ be an economy satisfying Assumption~U, then for each $i$, $j$ in $I$, the initial endowment $e^j$ is strongly desirable for agent $i$ in the sense that
\[
\forall x \in E_+, \quad \forall t>0, \quad V^i(x + te^j) > V^i(x).
\]
\end{remark}

\begin{remark}\label{rem:U1}
Assume that Assumption~U.2 is satisfied,
\begin{description}
\item[(a)] if preferences of agent $i$ are strictly increasing and $e^j>0$ for each $j\in J$, then Assumption~U.1 is satisfied,
\item[(b)] if preferences of agent $i$ are increasing and for each $j\in I$, there exists a strictly positive integrable adapted process $\xi^j$ such that $de^j(t)=\xi^j(t) dt$, then Assumption~U.1 is satisfied.
\end{description}
\end{remark}

In order to compare norm properness and the existence of smooth sub-gradients in $K$, we consider the following monotonicity assumption.
\begin{assumption}[M]
For every Edgeworth equilibrium $\bm{x}$, for each agent $i$, the following property is satisfied:
\[
\forall j\in I, \quad \forall t>0, \quad x^i+ t e^j + E_+ \subset P^i(x^i).
\]
\end{assumption}

\begin{remark}
From Remarks~\ref{rem:increasing}-\ref{rem:desirable}, Assumptions~C and Conditions~U.1 and U.2 imply Assumption~M.
\end{remark}

It is proved in \cite{Martinsda} that under Assumption~C.2, the existence of smooth sub-gradients in $K$ implies that the economy is weakly proper,\footnote{I.e., proper for the weak topology $\sigma(E,K)$.} in particular it is norm proper.

It was left as open question in \cite{HindyHuang92} whether a norm proper economy admits a continuous equilibrium price.
The existence results available in the literature are not general enough to be applied directly to our framework.
The topology derived from the intertemporal norm does not give rise to the mathematical properties known to be sufficient for the existence of an Arrow--Debreu equilibrium.\footnote{The topological dual space $(E,\norm{\cdot})'$ endowed with dual order defined by the cone $E^\star_+$ is not a vector lattice.}
The main contribution of this section is to prove that if an economy is proper with respect to the intertemporal norm it admits a compatible equilibrium, i.e., a price functional that is continuous on the positive cone $E_+$.

\begin{theorem}\label{th:ExistenceComp}
Under Assumptions~C and~M, if an economy is norm proper then it admits a compatible price.
\end{theorem}

\begin{proof}
Consider an economy satisfying Assumptions~C and~M and assume that it is norm proper.
From the norm properness of utility functions, there exists a family $(\psi^i)_{i\in I}$ where $\psi^i$ belongs to $K$ and supports agent $i$'s preferences.
In order to apply Proposition~2 and Theorem~2 in \cite{Martinsda}, it is sufficient to prove that the maximum of two processes in $K$ is a process in $H$.
Actually this is a consequence of the fact that $H$ is stable by taking the max.
Indeed, let $\phi$ and $\psi$ be two processes in $H$, i.e., $\phi$ and $\psi$ are nonnegative, rightcontinuous with left limits, bounded in $L^q$,  and the projection of a raw continuous process bounded in $L^1$.
We denote by $\theta$ the process defined by $\theta_t=\max\{\phi_t,\psi_t\}$.
We have to show that $\theta$ belongs to $H_+$.
It is nonnegative, rightcontinuous with left limits and bounded in $L^q$.
It remains to show that $\theta$ is the optional projection of a raw continuous process in $L^1$.
For this we can again check the conditions of the main result in \cite{Bismut78}.
To this end we have to show that $\theta$ is of class (D) and continuous in expectations.
As $\theta$ is bounded in $L^q$, it is of class (D).
Continuity in expectation is preserved by taking the max, and the proof is done.
\end{proof}

\begin{remark}
Actually, Theorem~\ref{th:ExistenceComp} is still valid if the norm properness of each utility function $V^i$ is replaced by the $\tau$-properness for any linear topology $\tau$ on $E$ such that any  linear functional $\tau$-continuous on $E$ is represented by a vector in $H$.
\end{remark}

\begin{remark}
Observe that contrary to \cite{BankRiedel02} and \cite{Martinsda}, we don't need to assume that the filtration $\mathbb{F}$ is quasi left-continuous.
This is an assumption on the way new information is revealed to the agents.
Economically, an information flow corresponds to a quasi left-continuous filtration\footnote{See \cite{HindyHuang92} for a precise definition. An information flow generated by a Brownian motion or a Poisson process is quasi left-continuous.} if information surprises (in the sense of \cite{HindyHuang92}) occur only at times which cannot be predicted.
The announcement of a policy change of the Federal reserve is an example for an information surprise which occurs at a time known in advance.
\end{remark}

\subsection{Example}

We consider {\it Hindy--Huang--Kreps preferences}, i.e. preferences given by utility functionals of the form
\[
V^i(x) = \E \int_{[0,T]} u^i(t,Y(x)(t)) \kappa(dt)
\]
where $u^i : [0,T] \times \R_+ \rightarrow \R$ denotes a felicity function for agent $i$, and the quantity
\[
Y(x)(t) =  \int_{[0,t]} \beta e^{-\beta(t-s)} dx(s)
\]
describes the investor's level of satisfaction obtained from his consumption up to time $t\in [0,T]$.
The constant $\beta>0$ measures how fast satisfaction decays.

We consider the linear mapping $\phi : E \rightarrow E$ defined by
\[
\forall t\in [0,T], \quad \phi(x)(t) = \int_{[0,t]} \exp\{\beta s\} dx(s).
\]
For each $x\in E$, the vector $\phi(x)$ is defined by the optional random measure $d[\phi(x)](t) = \exp\{\beta t\} dx(t)$.
The linear mapping $\phi$ is bijective and the inverse mapping $\phi^{-1}$ is given by
\[
\forall t\in [0,T], \quad \phi^{-1}(x)(t) = \int_{[0,t]} \exp\{-\beta s\} dx(s).
\]
We introduce on $E$ the following norm $\rho$:
\[
\forall x\in E, \quad \rho(x) := \norm{\phi(x)} = \E \int_{[0,T]} |\phi(x)(t)| \kappa(dt).
\]

It is proved in \cite[Lemma~2]{Martinsda} that the norm-topology and the $\rho$-topology coincide on $E_+$, and that the $\rho$-topological dual $(E,\rho)'$ coincides with the norm-topological dual $(E,\norm{.})'$.
In order to apply Theorem~\ref{th:ExistenceComp} it is sufficient to prove that $V^i$ is $\rho$-proper.
From \cite[Theorem]{Martinsda} this is a consequence of the following conditions: for each $i\in I$,
\begin{description}
\item[(V.1)] for each $t\in [0,T]$, the function $u^i(t,.) : \R_+ \rightarrow \R$ is continuous, strictly increasing and concave,
\item[(V.2)] for each $y\in \R_+$, the function $u^i(.,y) : [0,T] \rightarrow \R$ is $\mathcal{B}$-measurable and the function $u^i(.,0)$ belongs to $L^1(\mathcal{B},\kappa)$,
\item[(V.3)] for each $t\in [0,T]$ the right-derivative $\partial_y u^i(t,0+)$ exists and the function $\partial_y u^i(.,0+)$ belongs to $L^\infty_+(\mathcal{B},\kappa)$.
\end{description}

\section{Conclusion}

We show how the economically sensible intertemporal topology introduced by \cite{HindyHuang92} allows to derive general structural results about equilibrium state prices.
Using the  \emph{th\'{e}orie g\'{e}n\'{e}rale} of stochastic processes, we show that price functionals that are continuous on the consumption set can be represented by state prices with right-continuous sample paths that admit left limits.
Moreover, the state price is the optional projection of a process with continuous sample paths that is not necessarily adapted.


\bibliographystyle{kluwer}

\end{document}